# Low-Complexity Iterative Algorithms for (Discrete) Compressed Sensing

Robert F.H. Fischer, Susanne Sparrer
Institut für Nachrichtentechnik, Universität Ulm, Ulm, Germany,
Email: {robert.fischer, susanne.sparrer}@uni-ulm.de

*Abstract*—We consider iterative ("turbo") algorithms for compressed sensing. First, a unified exposition of the different approaches available in the literature is given, thereby enlightening the general principles and main differences. In particular we discuss i) the estimation step (matched filter vs. optimum MMSE estimator), ii) the unbiasing operation (implicitly or explicitly done and equivalent to the calculation of extrinsic information), and iii) thresholding vs. the calculation of soft values. Based on these insights we propose a low-complexity but well-performing variant utilizing a Krylov space approximation of the optimum linear MMSE estimator. The derivations are valid for any probability density of the signal vector. However, numerical results are shown for the discrete case. The novel algorithms shows very good performance and even slightly faster convergence compared to approximative message passing.

## I. INTRODUCTION

In *compressed sensing* (CS) a sparse vector has to be recovered from an underdetermined system of linear equations [7]. Specifically, we consider the vector input/output relation[1]

$$\boldsymbol{y} = \boldsymbol{A}\boldsymbol{x} + \boldsymbol{n} \,, \tag{1}$$

where $\boldsymbol{x} \in \mathcal{X}^L$ is sparse ($s \ll L$ non-zero entries), $\boldsymbol{A} \in \mathbb{R}^{K \times L}$, with $K < L$, is the sensing matrix, $\boldsymbol{n} \in \mathbb{R}^K$ is zero-mean i.i.d. Gaussian noise with variance $\sigma_N^2$ per component and independent of $\boldsymbol{x}$, and $\boldsymbol{y} \in \mathbb{R}^K$ is the observation.

Throughout the letter we assume to know the sparsity $s$ and to have some prior knowledge on $\boldsymbol{x}$ in form of the probability density function (pdf). Classically, the non-zero elements of $\boldsymbol{x}$ are assumed to be real-valued, i.e., $x_i \in \mathcal{X} = \mathbb{R}$. Often, the pdf is assumed to be $f_X(x) = \frac{L-s}{L}\delta(x) + \frac{s}{L}\mathcal{N}(0,1)$. In communication scenarios, the elements of $\boldsymbol{x}$ are typically drawn from a finite set (discrete CS), i.e., $x_i \in \mathcal{X} = \{x^{(1)} = 0, x^{(2)}, \ldots, x^{(|\mathcal{X}|)}\}$, see, e.g., [22]. Here, the pdf reads $f_X(x) = \sum_{j=1}^{|\mathcal{X}|} p^{(j)} \delta(x - x^{(j)})$, where $p^{(j)} = \Pr\{x^{(j)}\}$ are the probabilities.

The main task in CS is to recover $\boldsymbol{x}$ from $\boldsymbol{y}$, which can be stated as the following optimization problem ($s$ known)

$$\hat{\boldsymbol{x}} = \underset{\tilde{\boldsymbol{x}} \in \mathcal{X}^L}{\operatorname{argmin}} ||\boldsymbol{y} - \boldsymbol{A}\tilde{\boldsymbol{x}}||_2^2 \quad \text{s.t.} \quad ||\tilde{\boldsymbol{x}}||_0 = s \,. \tag{2}$$

This work was supported in parts by the Deutsche Forschungsgemeinschaft (DFG) under grant FI 982/8-1.

[1]Notation: $||\cdot||_p$ denotes the $\ell_p$ norm. $\boldsymbol{A}_{(l,m)} = A_{l,m}$ is the element in the $l^{\text{th}}$ row and $m^{\text{th}}$ column of $\boldsymbol{A}$. $\boldsymbol{a}_m$, $\bar{\boldsymbol{a}}_l$: $m^{\text{th}}$ column and $l^{\text{th}}$ row of $\boldsymbol{A}$, respectively. $\boldsymbol{A}^{\mathsf{T}}$ and $\boldsymbol{A}^{-1}$ denote the transpose and the inverse of $\boldsymbol{A}$, respectively. $\operatorname{diag}([a_1, \ldots, a_L])$ denotes a diagonal matrix with the given elements. $\boldsymbol{I}$ is the identity matrix. $\operatorname{E}\{\cdot\}$: expectation operator. $f_X(x)$: probability density function of random variable $X$. $\mathcal{N}(m,v)$: Gaussian distribution with mean $m$ and variance $v$. $\delta(x)$: Dirac delta. $M_A$, $M_H$: arithmetic and harmonic mean, respectively.

Meanwhile, there are plenty of algorithms solving this problem; a non-exhaustive list includes *orthogonal matching pursuit* (OMP) [20], *iterative hard thresholding* (IHT) [3], *iterative soft thresholding (IST)* [5], *(Bayesian) approximate message passing* ((B)AMP) [8], [2] or *turbo signal recovery* (TSR) [17], [18]. Many of these algorithms can easily be adapted to the discrete setting, cf., e.g., [22], [23] or the philosophy of model-based CS [1].

In this letter, we concentrate on iterative ("turbo") algorithms and give a unified exposition enlightening the general principles and main differences. Based on these insights we propose a low-complexity but well-performing variant utilizing a Krylov space approximation of the optimum estimator. All derivations are equally valid for the continuous as well as discrete case; numerical results are shown for discrete CS.

The letter is organized as follows. In Sec. II, iterative recovery algorithms are briefly reviewed and the main components are enlightened. An approximation of the optimum estimator is derived in Sec. III. Sec. IV presents numerical results on the recovery performance and the numerical complexity of the respective algorithms. The letter closes with brief conclusions in Sec. V.

## II. ITERATIVE RECOVERY ALGORITHMS AND THE OPTIMAL ESTIMATOR

In all fields of communication and signal processing, iterative (turbo) schemes have been proven to be very powerful [14]. Basically, iterative recovery algorithms for CS iterate over two main steps until some stopping criterion is met. First, we call it "Step E", an estimate (or proxy) $\tilde{\boldsymbol{x}}$ on the vector is calculated via a linear estimator $\boldsymbol{H}$. In this joint processing step the $\ell_2$-norm of the error is minimized ignoring the sparsity. Second, in "Step S", the sparsity, or more generally, the a-priori knowledge (pdf $f_X$) on the signal is taken into account via a non-linear, element-wise processing resulting in the estimate $\hat{\boldsymbol{x}}$. The basic structure of the recovery algorithm is shown in Alg. 1.

In addition to the estimates, their quality (reliability information) may be quantified. Variances $\sigma_E^2$ (estimation step) and $\sigma_S^2$ (sparsity/soft value), respectively, may be calculated and passed to the next step.

Finally, in the case of discrete compressed sensing, a final quantization to the elements in the signal set $\mathcal{X}$ has to be performed (not shown in Alg. 1), cf. [22].

**Alg. 1** $\hat{\bm{x}} = \text{recover}\left(\bm{y}, \bm{A}, \sigma_N^2, s, \text{f}_X\right)$

1: $\hat{\bm{x}} = \bm{0}, \ \bm{r} = \bm{y}, \ \sigma_S^2 = \frac{s}{L}$ // Init
2: **while** stopping criterion not met {
3:     calculate $\bm{H} = \text{fkt}(\bm{A}, \sigma_N^2, \sigma_S^2)$
4:     $\tilde{\bm{x}} = \hat{\bm{x}} + \bm{H}\bm{r}$, calculate $\sigma_E^2$ // Step E
5:     calculate $\text{T} = \text{fkt}(\text{f}_X, s, \sigma_E^2)$
6:     $\hat{x}_l = \text{T}(\tilde{x}_l), \ \forall l$, calculate $\sigma_S^2$ // Step S
7:     $\bm{r} = \bm{y} - \bm{A}\hat{\bm{x}}$
8: }

### A. Iterative Recovery Algorithms

*1) Iterative Hard/Soft Thresholding:* Meanwhile, the IHT [3] and IST [5] algorithm belong to the standard repertoire of compressed sensing. Here, the signal estimation step is done by choosing

$$\bm{H} = \bm{C}\bm{A}^\mathsf{T} \ , \qquad (3)$$

where $\bm{C} = \text{diag}(c_1, \ldots, c_L)$, with $c_l = 1/\|\bm{a}_l\|_2^2$ ($\bm{a}_l$: $l^\text{th}$ column of $\bm{A}$). In term of communications, this can be interpreted as applying the *matched filter* ($\bm{A}^\mathsf{T}$) and normalizing ($\bm{C}$) the main-diagonal elements of the end-to-end cascade $\bm{C}\bm{A}^\mathsf{T}\bm{A}$ to one. Since $\bm{H}$ is fix over the iterations, Line 3 of Alg. 1 is not active.

In the thresholding step, a non-linear function $\text{T}(\tilde{x})$ is applied symbol-wise to the estimate $\tilde{\bm{x}}$ of the estimation step. Basically, all elements except the $s$ elements with largest magnitude are set to zero;[2] for details see [3], [5]. In IHT and IST no reliability information ($\sigma_E^2$, $\sigma_S^2$) is calculated or utilized.

*2) Iterative Soft Feedback:* In [24] an improved version of IHT/IST, called *iterative soft feedback* (ISF), has been introduced. Main difference is that the simple magnitude-oriented thresholding step is replaced by the calculation of the so-called *soft values*, which minimize the expected squared error to the true value [27], mathematically[3]

$$\text{T}(\tilde{x}) = \text{E}\{x \mid \tilde{x}\} = \int_{-\infty}^{\infty} x \text{f}_X(x \mid \tilde{x}) \, \text{d}x \ . \qquad (4)$$

To this end, the variance $\sigma_E^2$ of the estimation step and the variance $\sigma_S^2$ of the soft values have to be tracked during the iterations. Using the matched filter as in IHT/IST, the estimation variance can be approximated for normalized i.i.d. Gaussian sensing matrices $\bm{A}$ as [24]

$$\sigma_E^2 = \frac{L}{K}\sigma_S^2 + \sigma_N^2 \ . \qquad (5)$$

The variance of the (biased) soft values $\hat{\bm{x}}_\text{B} = \text{T}(\tilde{\bm{x}})$ is given by $\sigma_{S,\text{B}}^2 = \frac{1}{L}\sum_{l=1}^L \sigma_\text{T}^2(\tilde{x}_l)$ with ($\text{T}'(\tilde{x})$: derivative of T w.r.t. $\tilde{x}$)

$$\sigma_\text{T}^2(\tilde{x}) = \sigma_E^2 \cdot \text{T}'(\tilde{x}) = \text{E}\{(x - \text{T}(\tilde{x}))^2 \mid \tilde{x}\} \ . \qquad (6)$$

---

[2]The thresholding function $\text{T}(\tilde{x})$ itself depends on the entire vector $\tilde{\bm{x}}$; the threshold has to be adjusted to the current vector (calculation of T in Line 5 of Alg. 1). Once the function is determined, it is applied element-wise (Line 6 of Alg. 1).

[3]Examples for this function can be found, e.g., in [24].

For optimum performance, the *bias*, inherent in any MMSE solution [15], [11], has to be removed. This operation can also be interpreted as the calculation of *extrinsic information* [26]. Given $\hat{\bm{x}}_\text{B} = \text{T}(\tilde{\bm{x}})$ and $\sigma_{S,\text{B}}^2$, we obtain the unbiased versions as [13], [26]

$$\sigma_S^2 = \left(\tfrac{1}{\sigma_{S,\text{B}}^2} - \tfrac{1}{\sigma_E^2}\right)^{-1} , \ \hat{\bm{x}} = \sigma_S^2 \cdot \left(\tfrac{\hat{\bm{x}}_\text{B}}{\sigma_{S,\text{B}}^2} - \tfrac{\tilde{\bm{x}}}{\sigma_E^2}\right) \ . \qquad (7)$$

*3) Approximate Message Passing:* In (B)AMP the same estimation matrix (matched filter) as in IHT/IST and ISF is used and, as in ISF, the soft values according to (4) are calculated. However, the unbiasing step is replaced by a different calculation of the residual $\bm{r}$, specifically $\bm{r} = \bm{y} - \bm{A}\hat{\bm{x}}_\text{B} + b\bm{r}$, where $b = \frac{L}{K}\frac{\sigma_{S,\text{B}}^2}{\sigma_E^2} = \frac{1}{K}\sum_{l=1}^L \text{T}'(\tilde{x}_l)$.

*4) Turbo Signal Recovery:* An iterative algorithm for the estimation of (complex-valued) sparse vectors, denoted as *turbo signal recovery (TSR)*, has been proposed in [17], [18]. The original work utilizes a matched-filter frontend and is limited to sensing matrices with orthonormal rows. For this setting, TSR is equivalent to ISF; only the unbiasing in the estimation step differs slightly (see Sec. II-B). TSR can be viewed as turbo decoding/equalization [28].

*5) Iterative MMSE/Soft-Feedback Algorithm:* Meanwhile, TSR has been generalized to arbitrary sensing matrices and the exposition has been simplified in [25]. The essential improvement is the application of the actual MMSE estimator with correct removal of the bias. We denote this Turbo variant (employing MMSE estimator and soft feedback) as *TMS* [26].

Moreover, instead of tracking *average* variances $\sigma_E^2$ and $\sigma_S^2$, *individual* (per element) variances (collected in diagonal correlation matrices) may be calculated and utilized. This *iterative individual MMSE/soft-feedback* (IMS) algorithm [25] shows very good performance for discrete compressed sensing.

### B. Estimation, Linear MMSE Solution, and Unbiasing

We now have a closer look at the signal estimation step and the unbiasing of the MMSE solution. In general, an estimate on $\bm{x}$, having the prior knowledge $\hat{\bm{x}}$, can be calculated by

$$\tilde{\bm{x}} = \hat{\bm{x}} + \bm{H}(\bm{y} - \bm{A}\hat{\bm{x}}) \ , \qquad (8)$$

where $\bm{H}$ is the estimation matrix. Assuming the elements of the remaining signal $\bm{x} - \hat{\bm{x}}$ to be i.i.d., i.e., $\bm{\Phi}_{ss} = \text{E}\{(\bm{x} - \hat{\bm{x}})(\bm{x} - \hat{\bm{x}})^\mathsf{T}\} = \sigma_S^2 \bm{I}$, and i.i.d. zero-mean, white noise, i.e., $\bm{\Phi}_{nn} = \text{E}\{\bm{n}\bm{n}^\mathsf{T}\} = \sigma_N^2 \bm{I}$, the optimum *biased* linear MMSE estimate is obtained via the following estimation matrix [15]

$$\bm{H}_\text{B} = \bm{A}^\mathsf{T}\left(\bm{A}\bm{A}^\mathsf{T} + \tfrac{\sigma_N^2}{\sigma_S^2}\bm{I}\right)^{-1} \ . \qquad (9)$$

The unbiased version is given by $\bm{H}_\text{U} = \bm{C}\bm{H}_\text{B}$, where the diagonal matrix $\bm{C} = \text{diag}(c_1, \ldots, c_L)$ is chosen such that the main diagonal elements of $\bm{C}\bm{K} \stackrel{\text{def}}{=} \bm{C}\bm{H}_\text{B}\bm{A}$ are one, i.e., $c_l = 1/K_{l,l}$, $l = 1, \ldots, L$.

The main diagonal elements of the covariance matrix of the error $\bm{e} = \tilde{\bm{x}} - \bm{x}$ can be written for the biased and unbiased MMSE solution, respectively, as [11, Footnote 5], [10]

$$\left[\bm{\Phi}_{ee,\text{B}}\right]_{l,l} = \sigma_S^2(1 - K_{l,l}) \ , \ \left[\bm{\Phi}_{ee,\text{U}}\right]_{l,l} = \sigma_S^2\frac{1 - K_{l,l}}{K_{l,l}} \ . (10)$$

In each case, for optimal performance, the bias in the estimate has to be removed and the average estimation variance has to be calculated. To this end, two different approaches are possible: Variant "AU" utilizes the biased MMSE solution, calculates the average ("A") variance of the biased solution, and then converts (via (7) but exchanging the meaning on "E" and "S") estimates and variance to the unbiased ("U") quantities. In [26] it is shown that this calculation of extrinsic quantities (cf. [13]) is identical to a joint unbiasing. Alternatively, in Variant "UA" the unbiased ("U") MMSE solution is directly calculated, hence individual unbiasing is performed. Then, the average ("A") variance is calculated.

Note, the average variance is given for Variant "UA" after straightforward manipulations as ($M_H$: harmonic mean)

$$\sigma_E^2 = \tfrac{1}{L}\operatorname{trace}(\boldsymbol{\Phi}_{ee,\mathrm{U}}) = \sigma_S^2 \cdot \left(\tfrac{1}{M_H(K_{l,l})} - 1\right) . \tag{11}$$

For variant "AU" the biased version first calculates to $\sigma_{E,\mathrm{B}}^2 = \tfrac{1}{L}\operatorname{trace}(\boldsymbol{\Phi}_{ee,\mathrm{B}})$, which, using the unbiasing formula (7) (cf. also [13], [26]), results after some manipulations in ($M_A$: arithmetic mean)

$$\sigma_E^2 = \left(\tfrac{1}{\sigma_{E,\mathrm{B}}^2} - \tfrac{1}{\sigma_S^2}\right)^{-1} = \sigma_S^2 \cdot \left(\tfrac{1}{M_A(K_{l,l})} - 1\right) . \tag{12}$$

Noteworthy, the two unbiasing strategies (averaging–unbiasing vs. unbiasing–averaging) are not identical. Since the elements $K_{l,l}$ of the end-to-end cascade are all positive and real, the relation $M_A(K_{l,l}) > M_H(K_{l,l})$ holds, which means that $\sigma_E^2$ will be larger for Variant "UA" than for Variant "AU". However, as long as the elements $K_{l,l}$ do not vary too much, arithmetic and harmonic mean almost do not differ and the performance will be almost identical.

## III. KRYLOV SPACE APPROXIMATION OF THE ESTIMATION MATRIX

The calculation of the estimation matrix $\boldsymbol{H}$ for the MMSE solution in (9) requires the inversion of a $K \times K$ matrix and hence has large numerical complexity, especially if it has to be done per iteration in the reconstruction algorithm.

Considering the series expansion [12]

$$(\boldsymbol{Q} + \boldsymbol{I})^{-1} = \sum_{i=0}^{\infty}(-\boldsymbol{Q})^i = \boldsymbol{I} - \boldsymbol{Q} + \boldsymbol{Q}^2 \pm \cdots , \tag{13}$$

which converges if the largest eigenvalue of $\boldsymbol{Q}$, $\lambda_{\max}(\boldsymbol{Q})$, is smaller than one, an approximation of the MMSE solution can be given by truncating this expansion. Let $\boldsymbol{Q} = \alpha\sigma_S^2 \boldsymbol{A}\boldsymbol{A}^\mathsf{T} + (\alpha\sigma_N^2 - 1)\boldsymbol{I}$, then the biased MMSE estimation matrix (9) can be written as [21]

$$\boldsymbol{H}_\mathsf{B} = \alpha\sigma_S^2 \boldsymbol{A}^\mathsf{T}\left(\boldsymbol{Q} + \boldsymbol{I}\right)^{-1} . \tag{14}$$

Via the choice of $\alpha$ the eigenvalues of $\boldsymbol{Q}$ and thus the convergence of the series in (13) can be controlled; for stability [21] $\alpha < \alpha_{\max} \stackrel{\text{def}}{=} 2/(\sigma_S^2 \lambda_{\max}(\boldsymbol{A}\boldsymbol{A}^\mathsf{T}) + \sigma_N^2)$ is required.

Please note that this type of approximation is well-known from the field of *code-division multiple access* (CDMA) schemes, e.g., [19], [16], [21]. Since the equalization result can be written as a linear combination of the receive vector $\boldsymbol{y}$ and $\boldsymbol{Q}\boldsymbol{y}, \boldsymbol{Q}^2\boldsymbol{y}, \ldots$, it can be interpreted as an approximation of $\boldsymbol{H}$ in a *Krylov space*, whose basis vectors are $\boldsymbol{Q}^i\boldsymbol{y}$, $i = 0, 1, \ldots$ [6].

### A. Zeroth-Order Approximation

The simplest approximation to (14) is

$$\boldsymbol{H}_\mathsf{B} = \alpha\sigma_S^2 \boldsymbol{A}^\mathsf{T} . \tag{15}$$

For the unbiased solution, the elements of the normalization matrix $\boldsymbol{C}$ are then given by $c_{l,l} = 1/(\alpha\sigma_S^2 \|\boldsymbol{a}_l\|_2^2)$, and the conventional proxy calculation via $\boldsymbol{A}^\mathsf{T}$, i.e., the matched filter, as in IHT/IST/ISF/AMP results. The estimation variance in case of i.i.d. Gaussian sensing matrices has been given in (5).

### B. First-Order Approximation

The second simplest approximation is

$$\boldsymbol{H}_\mathsf{B} = \alpha\sigma_S^2 \boldsymbol{A}^\mathsf{T}(\boldsymbol{I} - \boldsymbol{Q}) = \gamma \boldsymbol{A}^\mathsf{T}(\boldsymbol{I} - \beta\boldsymbol{A}\boldsymbol{A}^\mathsf{T}) , \tag{16}$$

with $\beta = \sigma_S^2/(2/\alpha - \sigma_N^2)$ and $\gamma = \alpha\sigma_S^2 \cdot (2 - \alpha\sigma_N^2)$. Please note that the choice of $\beta$ influences the "strength" of the first-order term $\boldsymbol{A}^\mathsf{T}\boldsymbol{A}$. For $\alpha < \alpha_{\max}$ we have $\beta < 1/\lambda_{\max}(\boldsymbol{A}\boldsymbol{A}^\mathsf{T})$.

The covariance matrix of the error $\boldsymbol{e}$ after unbiasing is then given by

$$\boldsymbol{\Phi}_{ee} = \sigma_S^2(\boldsymbol{H}\boldsymbol{A}\boldsymbol{A}^\mathsf{T}\boldsymbol{H}^\mathsf{T} - \boldsymbol{A}^\mathsf{T}\boldsymbol{H}^\mathsf{T} - \boldsymbol{H}\boldsymbol{A} + \boldsymbol{I}) + \sigma_N^2 \boldsymbol{H}\boldsymbol{H}^\mathsf{T} \\ = \sigma_S^2 \boldsymbol{M}_S + \sigma_N^2 \boldsymbol{M}_N , \tag{17}$$

with the obvious definition of $\boldsymbol{M}_S$ and $\boldsymbol{M}_N$. The average error variance $\sigma_E^2$ then calculates to

$$\sigma_E^2 = \tfrac{1}{L}\operatorname{trace}(\boldsymbol{\Phi}_{ee}) = \sigma_S^2 \mu_S + \sigma_N^2 \mu_N , \tag{18}$$

where $\mu_S = \tfrac{1}{L}\operatorname{trace}(\boldsymbol{M}_S)$ and $\mu_N = \tfrac{1}{L}\operatorname{trace}(\boldsymbol{M}_N)$.

Having selected a suited[4] $\beta$, $\boldsymbol{H}$ is fixed and $\mu_S$, $\mu_N$ can be calculated in advance. None of these parameters have to be recalculated during the iterations, which lowers the complexity significantly compared to TMS/IMS where the MMSE estimator has to be recalculated per iteration.

In Alg. 2 the recovery algorithm <u>i</u>terating over the first-order <u>K</u>rylov space approximation of the MMSE estimator and calculating <u>s</u>oft values, denoted as *IKS*, is given.

---

**Alg. 2** $\hat{\boldsymbol{x}} = \mathrm{IKS}\left(\boldsymbol{y}, \boldsymbol{A}, \boldsymbol{H}, \mu_S, \mu_N, \sigma_N^2, s, \mathsf{f}_X\right)$

1: $\hat{\boldsymbol{x}} = \boldsymbol{0}$, $\boldsymbol{r} = \boldsymbol{y}$, $\sigma_S^2 = \tfrac{s}{L}$     // Init
2: **while** stopping criterion not met {
3:    $\tilde{\boldsymbol{x}} = \hat{\boldsymbol{x}} + \boldsymbol{H}\boldsymbol{r}$, $\sigma_E^2 = \sigma_S^2\mu_S + \sigma_N^2\mu_N$    // Step E
4:    $\hat{x}_{l,\mathsf{B}} = \mathrm{T}(\tilde{x}_l)$, $\forall l$, $\sigma_{S,\mathsf{B}}^2 = \tfrac{1}{L}\sum_{l=1}^{L}\sigma_\mathrm{T}^2(\tilde{x}_l)$    // Step S
5:    calculate $\sigma_S^2$, $\hat{\boldsymbol{x}}$ acc. to (7)
6:    $\boldsymbol{r} = \boldsymbol{y} - \boldsymbol{A}\hat{\boldsymbol{x}}$
7: }

---

[4]This selection can either be done dependent on the current sensing matrix or on its statistical properties. For the former variant the (loose) estimate $\lambda_{\max,\mathrm{est}}(\boldsymbol{A}\boldsymbol{A}^\mathsf{T}) = \max_{l=1,\ldots,K}\sum_{k=1}^{K}|\bar{\boldsymbol{a}}_l \bar{\boldsymbol{a}}_k^\mathsf{T}|$, where $\bar{\boldsymbol{a}}_l$ denotes the $l^{\mathrm{th}}$ row of $\boldsymbol{A}$, can be used [21].

## IV. NUMERICAL RESULTS

We now present results from numerical simulations for $L = 258$ and $K = 129$; the sparsity $s$ is assumed to be known. The elements of the sensing matrix $\boldsymbol{A}$ are i.i.d. unit-variance zero-mean (real) Gaussian; the columns are normalized to unit $\ell_2$ norm. Discrete CS with $\mathcal{X} = \{0, +1, -1\}$ and $f_X(x) = \frac{s/2}{L}\delta(x+1) + \frac{L-s}{L}\delta(x) + \frac{s/2}{L}\delta(x-1)$ is considered. The algorithms either perform 20 iterations or stop if the squared Euclidean norm of the difference in $\hat{\boldsymbol{x}}$ between two iterations is less than $10^{-8}$. The results of the algorithms are finally quantized to the given set $\mathcal{X}$ guaranteeing the fixed sparsity.

In Fig. 1, the symbol error rate SER $= \frac{1}{L}\sum_{i=1}^{L} \Pr\{\hat{x}_i \neq x_i\}$ averaged over a large number of sensing matrices and signal vectors is displayed over the inverse noise level in dB. The sparsity is fixed to $s = 12$.

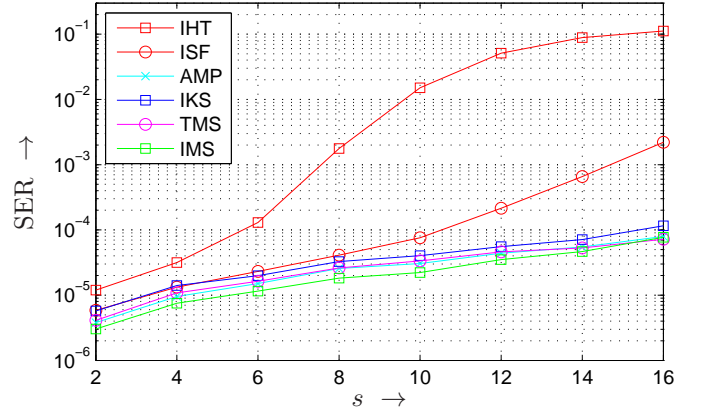

Fig. 2. SER over the sparsity. $L = 258$, $K = 129$, Gaussian i.i.d. sensing matrix with columns normalized to unit norm, $10\log_{10}(1/\sigma_N^2) = 17$ dB, $\mathcal{X} = \{0, +1, -1\}$.

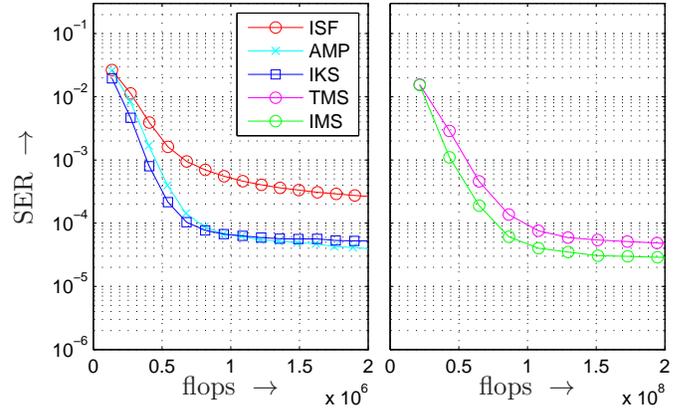

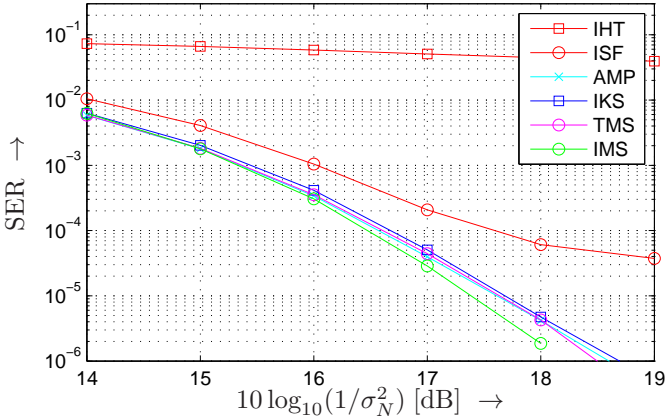

Fig. 1. SER over the noise level $1/\sigma_n^2$ in dB. $L = 258$, $K = 129$, Gaussian i.i.d. sensing matrix with columns normalized to unit norm, $s = 12$, $\mathcal{X} = \{0, +1, -1\}$.

Fig. 3. SER over the complexity (FLOPs). $L = 258$, $K = 129$, Gaussian i.i.d. sensing matrix with columns normalized to unit norm, $s = 12$, $10\log_{10}(1/\sigma_N^2) = 17$ dB, $\mathcal{X} = \{0, +1, -1\}$.

As can be seen, IHT does not provide satisfactory results. The curve of ISF flattens out at low SER. IMS with individual feedback shows the best performance. TMS, where average feedback is used, has only a slight disadvantage compared to IMS. AMP performs similar to TMS; IKS is almost as good as TMS/AMP.

Next, in Fig. 2 the average symbol error rate is plotted over the sparsity $s$. The inverse noise level is fixed to 17 dB.

IHT fails if the sparsity is larger than approx. 8; performance of ISF starts to degrade for sparsity larger than approx. 10. IMS, TMS, AMP, and IKS perform very similar with a slight advantage for IMS.

Finally, in Fig. 3 the evolution of the average symbol error rate during the iterations is plotted over the numerical complexity,[5] which is expressed as the number of floating-point operations (FLOPs, counted during the run of the algorithms). Each marker corresponds to one iteration. Please note the different scalings of the left and right part of Fig. 3.

ISF, AMP, and IKS have almost the same complexity in each iteration, dominated by the multiplication with the fixed $L \times K$ estimation matrix. IMS and TMS achieve good results after 7 or 8 iterations but each iteration (due to inverting a $K \times K$ matrix) has high complexity (approx. 100 times the complexity of AMP or IKS). IKS achieves almost the same SER as IMS and similar results as AMP. Over the first iterations, IKS has the best performance at the lowest complexity, i.e., a slightly faster convergence rate as AMP. If only 4 to 8 iterations are allowed in total, IKS is the best choice.

## V. CONCLUSIONS

We have given an overview on iterative algorithms for (discrete) compressed sensing, enlightening the main operations and differences. Starting from the optimum MMSE solution and discussing the unbiasing step, an approximation of the estimation matrix has been proposed. The resulting algorithm has a slightly better convergence compared to AMP at the same complexity.

---

[5]Assuming the sensing matrix $\boldsymbol{A}$ to change only very infrequently, the initial complexity for calculating $\boldsymbol{H}$ can be ignored and only the complexity of the iterations is relevant.

## References

[1] R.G. Baraniuk, V. Cevher, M.F. Duarte, C. Hedge. Model-Based Compressive Sensing. *IEEE Transactions on Information Theory*, vol. 56, no. 4, pp. 1982–2001, Apr. 2010.

[2] M. Bayati, A. Montanari. The Dynamics of Message Passing on Dense Graphs, with Application to Compressed Sensing. *IEEE Transactions on Information Theory*, vol. 57, no. 2, pp. 764–785, Feb. 2011.

[3] T. Blumensath, M.E. Davis. Iterative Thresholding for Sparse Approximations. *Journal of Fourier Analysis and Applications*, vol. 14, no. 5, pp. 629–654, Dec. 2008.

[4] J.M. Cioffi, G.P. Dudevoir, M.V. Eyuboglu, G.D. Forney. MMSE Decision-Feedback Equalizers and Coding. Part I; Equalization Results. *IEEE Transactions on Communications*, vol. 43, no. 10, pp. 2582–2594, Oct. 1995.

[5] I. Daubechies, M. Fornasier, I. Loris. Accelerated Projected Gradient Method for Linear Inverse Problems with Sparsity Constraints. *Journal of Fourier Analysis and Applications*, vol. 14, no. 8, pp. 764–792, Dec. 2008.

[6] G.K.E. Dietl. *Linear Estimation and Detection in Krylov Subspaces*. Springer-Verlag Berlin Heidelberg, 2007.

[7] D.L. Donoho. Compressed Sensing. *IEEE Transactions on Information Theory*, vol. 52, no. 4, pp. 1289–1306, Apr. 2006.

[8] D.L. Donoho, A. Maleki, A. Montanari. Message Passing Algorithms for Compressed Sensing: I. Motivation and Construction. In *Proc. Information Theory Workshop (ITW) 2010*, Cairo, Egypt, Jan. 2010.

[9] R.F.H. Fischer. MMSE DFE for High-Rate MIMO Transmission over Channels with ISI. In *Proceedings of the 5th IEE International Conference on 3G Mobile Communication Technologies 3G*, pp. 83–87, London, United Kingdom, Oct. 2004.

[10] R.F.H. Fischer, C. Siegl. Inflated Lattice Precoding, Bias Compensation, and the Uplink/Downlink Duality: The Connection. *IEEE Communications Letters*, vol. 11, no. 2, pp. 185–187, Feb. 2007.

[11] G.D. Forney Jr. On the role of MMSE estimation in approaching the information-theoretic limits of linear Gaussian channels: Shannon meets Wiener. *Proceedings of the 41st Allerton Conf. on Communication, Control, and Computing*, Monticello, IL, Oct. 2003.

[12] G. Golub, C. Van Loan. *Matrix Computations*. The Johns Hopkins University Press, Baltimore, 1996.

[13] Q. Guo, D.D. Huang. A Concise Representation for the Soft-in Soft-out LMMSE Detector. *IEEE Communications Letters*, vol. 15, no. 5, pp. 566–568, May 2011.

[14] J. Hagenauer. The Turbo Principle: Tutorial Introduction and State of the Art. In *Proc. International Symposium on Turbo Codes and Related Topics*, pp. 1–11, Brest, France, Sep. 1997.

[15] S.M. Kay. *Fundamentals of Statistical Signal Processing: I. Estimation Theory*, Prentice-Hall Inc., Upper Saddle River, NJ, USA, 1993.

[16] Z.D. Lei, T.J. Lim. Simplified Polynomial-Expansion Linear Detectors for DS-CDMA Systems. *Electronics Letters*, vol. 34, no. 16, pp. 1561–1563, Aug. 1998.

[17] J. Ma, X. Yuan, L. Ping. Turbo Compressed Sensing with Partial DFT Sensing Matrix. *IEEE Signal Processing Letters*, vol. 22, no. 2, pp. 158–161, Feb. 2015.

[18] J. Ma, X. Yuan, L. Ping. On the Performance of Turbo Signal Recovery with Partial DFT Sensing Matrices. *IEEE Signal Processing Letters*, vol. 22, no. 10, pp. 1580–1584, Oct. 2015.

[19] S. Moshavi, E.G. Kanterakis, D.L. Schilling. Multistage Linear Receivers for DS-CDMA Systems. *International Journal of Wireless Information Networks*, vol. 3, no. 1, pp. 1–17, Jan. 1996.

[20] Y.C. Pati, R. Rezaiifar, P.S. Krishnaprasad. Orthogonal Matching Pursuit: Recursive Function Approximation with Applications to Wavelet Decomposition. In *Proc. Asilomar Conf. on Signals, Systems and Computers*, pp. 40–44, Nov. 1993.

[21] G.M.A. Sessler, F.K. Jondral. Rapidly Converging Polynomial Expansion Multiuser Detector with low Complexity for CDMA Systems. *Electronics Letters*, vol. 38, no. 17, pp. 997–998, Aug. 2002.

[22] S. Sparrer, R.F.H. Fischer. Adapting Compressed Sensing Algorithms to Discrete Sparse Signals. In *Proc. Workshop on Smart Antennas (WSA) 2014*, Erlangen, Germany, Mar. 2014.

[23] S. Sparrer, R.F.H. Fischer. Soft-Feedback OMP for the Recovery of Discrete-Valued Sparse Signals. In *Proc. 23th European Signal Processing Conference (EUSIPCO'15)*, Nice, France, Aug. 2015.

[24] S. Sparrer, R.F.H. Fischer. Enhanced Iterative Hard Thresholding for the Estimation of Discrete-Valued Sparse Signals. In *Proc. 24th European Signal Processing Conference (EUSIPCO'16)*, Budapest, Hungary, Aug. 2016.

[25] S. Sparrer, R.F.H. Fischer. Algorithms for the Iterative Estimation of Discrete-Valued Sparse Vectors. In *Proc. 11th International ITG Conference on Systems, Communications and Coding (SCC)*, Hamburg, Germany, Feb. 2017.

[26] S. Sparrer, R.F.H. Fischer. Unveiling Bias Compensation in Turbo-Based Algorithms for (Discrete) Compressed Sensing. Preprint. Available at http://arxiv.org/abs/1703.00707, March 2017.

[27] F. Tarköy. MMSE-Optimal Feedback and its Applications. In *Proc. IEEE International Symposium on Information Theory (ISIT)*, p. 334, Whistler, British Columbia, Canada, Sep. 1995.

[28] M. Tüchler, A.C. Singer. Turbo Equalization: An Overview. *IEEE Transactions on Information Theory*, vol. 57, no. 2, pp. 920–952, Feb. 2011.